\documentclass[12pt,twoside]{JHEP3}

\usepackage{amsmath,graphicx}
\usepackage{cite}
\author{Stefan Gieseke\\PH Department, CERN, 1211 Geneva 23, Switzerland\\ 
  \emph{and} Institut f\"ur Theoretische Physik,\\ Universit\"at Karlsruhe, 
  76128 Karlsruhe, Germany.
}

\author{Michael H.~Seymour\\PH Department, CERN, 1211 Geneva 23, Switzerland\\ 
  \emph{and} School of Physics \& Astronomy, University of Manchester, U.K.}

\author{Andrzej Si\'odmok\\PH Department, CERN, 1211 Geneva 23, Switzerland\\ 
  \emph{and} Marian Smoluchowski Institute of Physics,\\ 
Jagiellonian University, ul. Reymonta 4, 30-059 Cracow, Poland.
}

\abstract{We consider a model of transverse momentum production in
  which non-perturbative smearing takes place throughout the
  perturbative evolution, by a simple modification to an initial state
  parton shower algorithm.  Using this as the important
  non-perturbative ingredient, we get a good fit to data over a wide
  range of energy.  Combining it with the non-perturbative masses and
  cutoffs that are a feature of conventional parton showers also leads
  to a reasonable fit.  We discuss the extrapolation to the LHC.}

\title{A model of non-perturbative gluon emission in an  
initial state parton shower}


\preprint{arXiv:0712.1199\\
CERN-PH-TH/2007-240\\
KA-TP-31-2007}

\newcommand{\hpp}{Herwig++}
\newcommand{\ResBos}{ResBos}

\begin{document}

\section{Introduction}

At LEP and the SLC, the properties of W and Z bosons could be studied to
great accuracy because to a very good approximation they could be
calculated using only electroweak perturbation theory.  At the current
Tevatron and future LHC colliders, on the other hand, the event rates
are enormous and the expected statistical precision excellent, but the
electroweak bosons are produced by the annihilation of coloured partons
that are initially confined into colourless hadrons.  This means that
QCD effects, both perturbative and non-perturbative, play an extremely
important role in determining the properties of events containing
electroweak bosons and the limited precision with which we can calculate
those effects will ultimately be responsible for the dominant systematic
uncertainties on the measurements.

In this paper we will concentrate on one particular property of the
produced W and Z bosons\footnote{We are also interested in virtual
photons with invariant masses well below that of the Z, particularly
for tuning and validating our model.  All our calculations include
properly the full interference between $\gamma^*$ and Z, but with an
eye on the ultimate application at the LHC, we continue to refer in
this introduction to Zs.}, namely their transverse momentum
distribution.  This is interesting from the QCD point of view as,
sweeping across the distribution, one has regions dominated by hard
perturbative emission, multiple soft and/or collinear, but still
perturbative, emission, and truly non-perturbative confinement effects.
It is also an important quantity for the experimental programme, because
the W reconstruction efficiency is transverse momentum dependent, having
a direct effect on the ultimate precision of the W mass measurement as
well as helping the understanding of the signature for Higgs boson
production at either the Tevatron or the LHC \cite{Balazs:2000sz}.
Although the experiments measure the Z transverse momentum distribution
and use this to infer that of the W, the extent to which the effects are
non-universal limits the ultimate accuracy of the measurement, unless
elaborate tricks as proposed in Ref.\cite{StandardCandle} are used.
Thus, a deeper theoretical understanding and more reliable models are
certainly needed.

The two approaches to predicting the transverse momentum distribution
are analytical resummation
\cite{Dokshitzer:1978yd,Altarelli:1978pn,Collins:1984kg,Davies:1984sp,Ellis:1997sc,Ellis:1997ii,RESBOS}
and parton shower algorithms
\cite{Sjostrand:1985xi,Marchesini:1987cf,Gieseke:2003rz} (there have
also been attempts to combine the two approaches \cite{Mrenna:1999mq}).
We will focus on the latter, but will draw a few comparisons with the
former later.  The parton shower approach starts from the tree-level
matrix element, usually supplemented by `matrix element corrections'
\cite{Corcella:1998rs,Seymour:1994df,Gieseke:2003rz,Catani:2001cc,Bengtsson:1986hr,Miu:1998ju}
that use higher-order tree-level matrix elements to describe emission at
scales of order the W/Z boson mass and higher\footnote{We do not go into
their details, but use the implementation of \cite{seyi} throughout
this paper.}.  These give a significant tail of events with very high
transverse momenta.  The hard events are then evolved down to low scales
by using the backward evolution parton shower approach
\cite{Sjostrand:1985xi}.  Recoil from the gluons
emitted\footnote{Together with other backward-evolution steps such as an
incoming sea quark being evolved back to an incoming gluon by emitting
a corresponding antiquark.} during this evolution build up a
transverse momentum for the W/Z.  The evolution terminates at some scale
of order the confinement or typical hadron mass scale.  However,
confinement effects, described for example as the Fermi motion of
partons within the hadron, mean that the partons initiating the shower
should have a non-perturbative transverse momentum distribution, often
described as their `intrinsic' transverse momentum, which is also
transferred to the W/Z by recoil \cite{Altarelli:1978pn}.

One way to implement the colour coherence inherent in QCD is to
formulate the parton shower as an evolution in (energy times) opening
angle, as implemented in the HERWIG \cite{Marchesini:1991ch} and \hpp
\cite{Gieseke:2003hm,herwig++,herwigman} event generators.  Analysis of higher
order corrections shows that the scale of the running coupling used in
this evolution should be of order the transverse momentum of the
emission \cite{Amati:1979fg,Catani:1990rr}, and once this is done one
must introduce an infrared cutoff in transverse momentum that is active
during every step of the evolution.  That is, the probability of each
backward step in the evolution variable, even at large values of that
variable, is logarithmically dependent on the cutoff.  In
Ref.~\cite{Borozan:2002fk}, one of us advocated the view that
conventional infrared cutoff scales on perturbative emission (in that
case on the transverse momenta used to describe the minijet production
in an underlying event model) should be thought of as infrared
\emph{matching} scales, with a non-perturbative model of emission below
the cutoff supplementing the usual perturbative one above.  In this
paper we propose such a model for backward evolution in which an
additional non-perturbative component at low transverse momentum
provides additional smearing at each step of the evolution.

We are particularly motivated by the fact that, in order to fit data,
conventional parton shower models need an `intrinsic' transverse
momentum $\langle k_T \rangle$ that grows with collision energy.
For example in Herwig++ its value grows from $\langle k_T \rangle=0.9$ GeV 
which is needed to describe the data taken at the energy $\sqrt{S}=62$ GeV 
(experiment R209) to $2.1$ GeV which is needed at the Tevatron energies 
($\sqrt{S}=1800$ GeV). One would expect
the average `intrinsic' transverse momentum per parton to be of the
order of $0.3-0.5$ GeV based solely on the proton size and uncertainty rule,
but the values extracted from data, even with attempts to reduce its
value \cite{Thome:2004sk} are too large and cannot be interpreted as ``intrinsic''.
As we shall see, different models of this energy dependence that fit current
data give very different predictions for the LHC.  In our model, this
growth is under some kind of `semi-perturbative' control, since the
amount of non-perturbative smearing grows with the length of the
perturbative evolution ladder.  We ask the question whether, with this
additional source of non-perturbative transverse momentum, a truly
intrinsic transverse momentum distribution for the initial partons, that
does not depend on the collision energy or type, is sufficient.
\FIGURE[t]{%
  \includegraphics[scale=0.9]{ptZ.8}\hfill
  \includegraphics[scale=0.9]{ptZ.9}
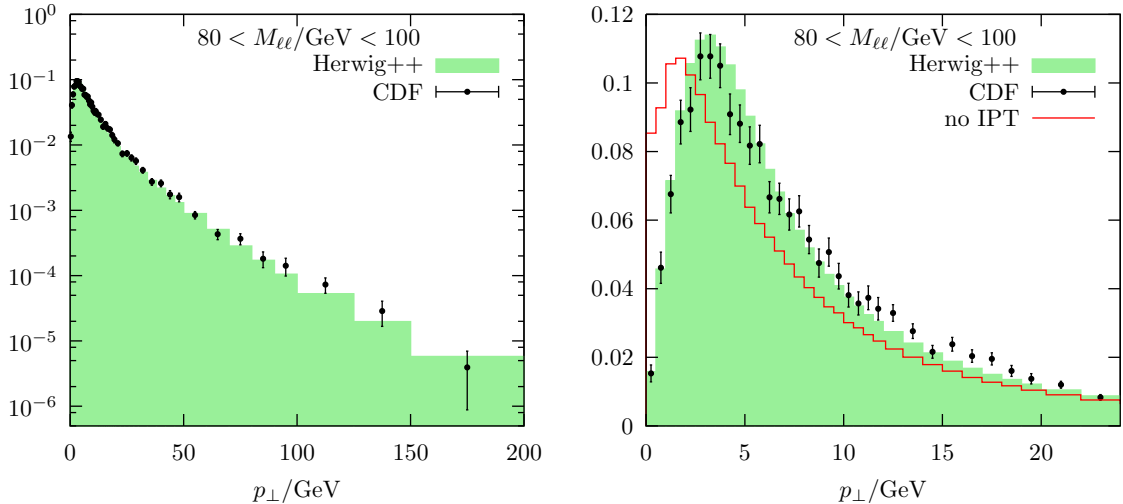
  \caption{The transverse momentum distribution of Z bosons at
    Tevatron energies compared to CDF data. Up to large transverse
    momenta (left) and only the small $p_\perp$ region (right).  The
    line denoted ``no IPT'' is from Herwig++ with intrinsic transverse
    momentum off. 
    \label{fig:tev-def} }
}

In Fig.~\ref{fig:tev-def} we show a comparison of the Z--boson
transverse momentum spectrum at Tevatron Run I with CDF data
\cite{CDFdata}.  The left panel shows that a description is possible
up to large transverse momentum.  The high transverse momentum region
is, however, dominated by contributions from hard gluon emissions.
These will not be the focus of this paper.  In general, the large
transverse momentum region will not be affected by soft,
non--perturbative emissions.

In the right panel of Fig.~\ref{fig:tev-def} we see only the small
transverse momentum region.  The Herwig++ result is shown with an
intrinsic $p_\perp=2.1\,$GeV from Gaussian smearing \cite{seyi}, which
is the default value at Tevatron energies.  To show the importance of
this effect we also plot the result with intrinsic $p_\perp$ set to
zero.  Clearly, this non--perturbative Gaussian smearing only affects
the region of small transverse momenta.  At large boson $p_\perp$ the
recoil against hard, perturbative gluon radiation dominates the
spectrum.

We also compared to D0 data \cite{D0data} and found a similar
agreement.  However, the CDF data has a finer binning and is therefore
more suitable for our comparison.

\section{Description of the model}
In order to simulate non--perturbative emission with the parton shower,
we consider the Sudakov form factor for backward evolution from some
scale $\tilde q_{\rm max}$ down to $\tilde q$ that is implemented in the
parton shower Monte Carlo program \hpp{}.  For further details, cf.\
Ref.~\cite{sudakovpaper}
\begin{equation}
  \label{eq:sudakov}
  \Delta(\tilde q; p_{\perp_{\rm max}},p_{\perp_0}) = \exp 
  \left\{-\int_{\tilde q^2}^{\tilde q^2_{\rm max}}
    \frac{d\tilde q'^2}{\tilde q'^2} \int_{z_0}^{z_1} 
    dz \frac{\alpha_S(p_{\perp})}{2\pi} 
    \frac{x'f_b(x', \tilde q'^2)}{xf_a(x, \tilde q'^2)}
  P_{ba}(z, \tilde q'^2)\right\} \ ,
\end{equation}
with $x'=x/z$.  The argument of the strong coupling $\alpha_S$ in
Eq.~(\ref{eq:sudakov}) is the transverse momentum $p_{\perp}$ of an
emission\footnote{Generally the scale of $\alpha_S$ is a function of the
evolution variables $z$ and $\tilde q^2$ and by default in Herwig++,
the argument of $\alpha_S$ is a slightly simplified expression, equal
to the transverse momentum to the required accuracy, but not exactly.
We have tested the implementation of our model with this simplified
expression and the exact expression for transverse momentum, and find
very similar results.  We therefore use the default expression.}.  The
cut-off scale at which the coupling would diverge, if extrapolated
outside the perturbative domain is represented by $p_{\perp_0}$.
Therefore two arguments of the Sudakov formfactor, $p_{\perp_{\rm max}}$ and
$p_{\perp_0}$ are not the evolution variables but only explicitly denote
the available phase-space of an emission.
  
We can introduce additional non--perturbative emissions in terms of an 
additional Sudakov form factor $\Delta_{NP}$, such that we have
\begin{equation}
  \label{eq:npsud}
  \Delta(\tilde q; p_{\perp_{\rm max}},0) 
  = \Delta_{\rm pert}(\tilde q; p_{\perp_{\rm max}},p_{\perp_0})
  \Delta_{\rm NP}(\tilde q; p_{\perp_0},0) 
\end{equation}

For technical simplicity we can achieve this by modifying our
implementation of $\alpha_S(p_{\perp})$ in such a way that we can extend
it into the non--perturbative region,
\begin{equation}
  \label{eq:asmod}
  \alpha_S(p_{\perp}) = \alpha_S^{(\rm pert)}(p_{\perp}) 
  + \alpha_S^{(\rm NP)}(p_{\perp}).  
\end{equation}
In our implementation we have not chosen $\alpha_S^{(\rm pert)}
(p_{\perp})$ and $\alpha_S^{(\rm NP)}(p_{\perp})$ explicitly but rather
modified the sum $\alpha_S(p_{\perp})$, in order to behave differently
in two physically different regions, divided by a separation scale
$p_{\perp_0}$,
\begin{equation}
  \label{eq:asdef}
  \alpha_S(p_{\perp}) =
  \begin{cases}
    \varphi(p_{\perp}), &p_{\perp}<p_{\perp_0}\\
    \alpha_S^{(\rm pert)} (p_{\perp}), &p_{\perp}\geq p_{\perp_0}
  \end{cases}
  \ .
\end{equation}
In this way, the kinematics and phase space of each non--perturbative
emission are exactly as in the perturbative case.  We only modify
their probabilities in the region of small transverse momenta. 

We have studied two simple choices of the non--perturbative function
$\varphi(p_{\perp})$ in greater detail:  
\begin{enumerate}
\item[(a)] ``\emph{flat}'': the flat continuation of $\alpha_S(p_{\perp}<p_{\perp_0})$
  with a constant value $\varphi_0 = \varphi(0)$,
  \begin{equation}
    \alpha_S(p_{\perp}<p_{\perp_0}) = \varphi_0 \ .
    \label{eq:asflat}
  \end{equation}  
\item[(b)] ``\emph{quadratic}'':  a quadratic interpolation between
  the two values $\alpha_S(p_{\perp_0})$ and $\varphi(0)$.  
  \begin{equation}
    \label{eq:asquad}
    \alpha_S(p_{\perp} < p_{\perp_0}) =  \varphi_0 +
    (\alpha_S(p_{\perp_0}) - \varphi_0) \frac{p_{\perp}^2}{p_{\perp_0}^2} \ .
  \end{equation}
\end{enumerate}
In both cases our model is determined by the two free parameters $p_{\perp_0}$
and $\varphi_0$.

We have concentrated our study on the small transverse momentum region
of vector boson production.  Therefore, the only modification of the
Herwig++ code that had to be made was the introduction of the two
non--perturbative parameters to $\alpha_S(p_{\perp})$.  In fact, as we
implemented it this would also affect final state radiation but our
observable is not sensitive to effects in the final state.  Details of
final state effects will be discussed in Sec.~\ref{sec:fsr}.

We would like to emphasise that we want to keep this model as simple
as possible in order to explore the possibility of a reasonable
description of the data.  Therefore, the shape of $\alpha_S$ in the
non--perturbative region is only a crude guess.  A further study of
the details of the shape would go beyong the scope of this work.

\section{Parton-level results}

To simulate fully exclusive events, Monte Carlo event generators like
Herwig++ use a hadronization model, which is assumed to be universal
across different types of collision and different processes within
them.  Therefore for our final results presented in sections
\ref{HadronLevel} and \ref{LHC} we will combine our model for
non-perturbative gluon emission with the standard Herwig++ model for
the termination of the shower using non-perturbative effective parton
masses tuned to $e^+e^-$ data so that the corresponding hadronization
model can be used.  However, if we are only interested in the W/Z
transverse momentum distribution, we do not need to hadronize the
final state: we can terminate the simulation at the end of the parton
shower.  We can therefore make a purely parton-level study with all
light quark and gluon effective masses and cutoffs set to
zero\footnote{For technical reasons it is not possible to set them
exactly to zero.  However, we have confirmed that if they are small
enough their precise values become irrelevant.  For this study we
actually set the quark masses and the $\delta$ parameter to 1~MeV, so
that the non-perturbative mass that cuts off the parton shower, called
$Q_g$ in the Herwig++ manual, is given by the cParameter.  For the
cParameter we ran with values in the range 10~MeV to 100~MeV and found
very little effect.  We therefore use 100~MeV for our main results.
% I have removed the comments about the running speed.  
} with our
model for the low-scale $\alpha_S$ as the only non-perturbative input.

The first observation that we can make with our model is that we can
easily find parameter values that describe existing Tevatron data.
However the main focus of
our work is on the understanding of the dependence of the
non--perturbative effects on the typical centre of mass (CM) energy of
the system or even the collider.  We therefore considered two more
datasets. The first is Fermilab E605 \cite{E605data} fixed target
$p$--Cu data, taken at 38.8\,GeV CM energy.  We only take the data with
an invariant mass of $11.5 < M/{\rm GeV} < 13.5$ as this goes out to the
highest transverse momentum.  The other data we consider were taken in
$p$--$p$ collisions at $\sqrt{S}=62\,$GeV at the CERN ISR experiment
R209 \cite{R209data}.  There are more data available but all at even
lower CM energies.  Our main interest is in finding a reasonable
extrapolation to LHC energies that is still compatible with the early
data.

We have run Herwig++ with varying non--perturbative parameters
$\varphi_0$ and $p_{\perp_0}$ for the two forms of $\alpha_S$ in
\eqref{eq:asflat} and \eqref{eq:asquad}.  After an initial broader
scan, we focussed on the region of $\varphi_0$ between 0 and 1 and
$p_{\perp_0}$ between 0.5\,GeV and 1.0\,GeV.  Each parameter set was
run for the three different experimental setups we consider.  We left
the intrinsic $k_\perp$ fixed at 0.4\,GeV.  For each resulting
histogram we have computed a total $\chi^2/$bin in order to quantify
its agreement or disagreement with the data.  We took the data errors
to be at least 5$\%$ as we did not want to bias towards exceptionally
good data points.  Furthermore, we ignored an additional systematic
error of the two fixed target data sets which is quoted to be around
5--10\%.  \FIGURE{%
  \includegraphics[scale=1]{chisq-part.2}
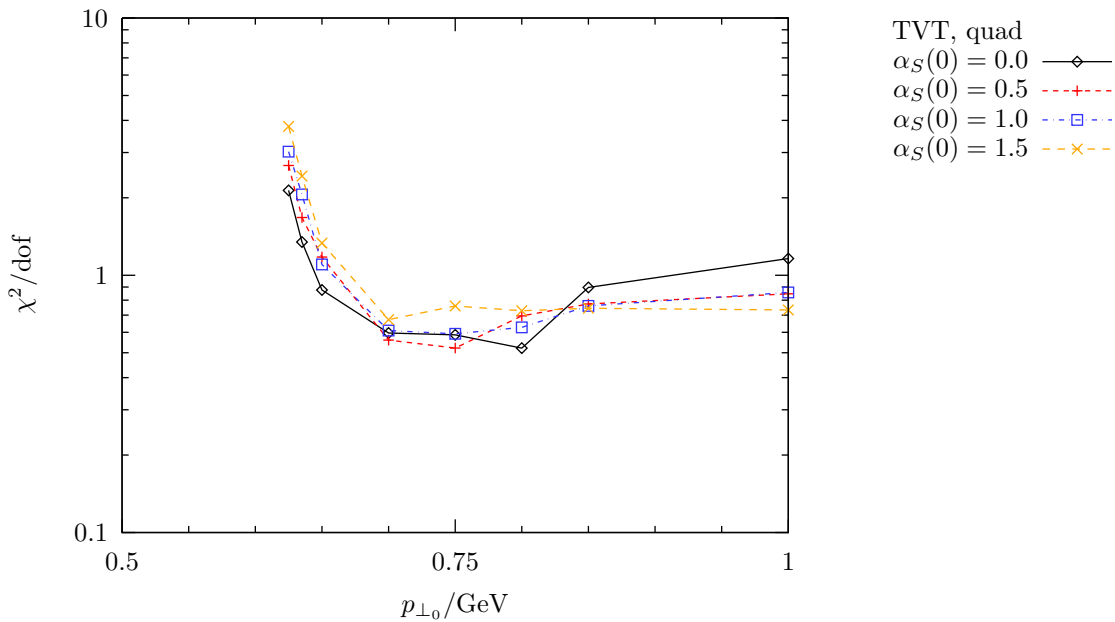
  \caption{$\chi^2$ values for the quadratic non--perturbative model
    compared to Tevatron data as a function of the NP parameter
    $p_{\perp_0}$.  The different lines are for different values of $\varphi_0
    = \alpha_S(0)$. 
    \label{fig:tev-chisq} } } 
Fig.~\ref{fig:tev-chisq} shows the
$\chi^2$ values we obtain for the quadratic model compared to Tevatron
data.  We made similar plots for the other two energies.  The basic
features are the same.  In each case we find clear minima within the
given $p_{\perp_0}$ range.  In going from one experiment to another we
find the more or less sharp minima.  The minimum in
Fig.~\ref{fig:tev-chisq} is not as clear as in the cases of the other
two experiments.  The best and most stable situation for all
experiments is found for $\alpha_S(0)=0.0$ and $p_{\perp_0} =
0.75\,$GeV.  The $\chi^2$ values are not very sensitive to the value
$\alpha_S(0)$ around the minimum, i.e.\ we are not very sensitive to
the non--perturbative region itself.  In Fig.~\ref{fig:alpha_s} we
show the non--perturbative region of our $\alpha_S$ parametrisation.
We have inspected all distributions directly as well and found a
consistency with this choice.  For this optimal choice over the energy
range 38.8\,GeV to 1.8\,TeV we show the resulting low $p_\perp$
distributions in Fig.~\ref{fig:bestpt-parton}.  We should stress that the
used parameter set may not be the optimal choice for each experiment
or CM energy but rather the best compromise between the three
experiments.  As the fixed target data do not even include the
systematic errors quoted we have deliberately put a bit more emphasis
on the Tevatron result.  Ultimately, our goal will be to extrapolate
our results further to LHC energies and we believe that for this
purpose we have made the right choice of parameters.  \FIGURE[h]{%
  \includegraphics[scale=0.7]{ptscan.0}\hspace{1cm}
  \includegraphics[scale=0.7]{ptscan.1}\\[5pt]
  \includegraphics[scale=0.7]{ptscan.2}
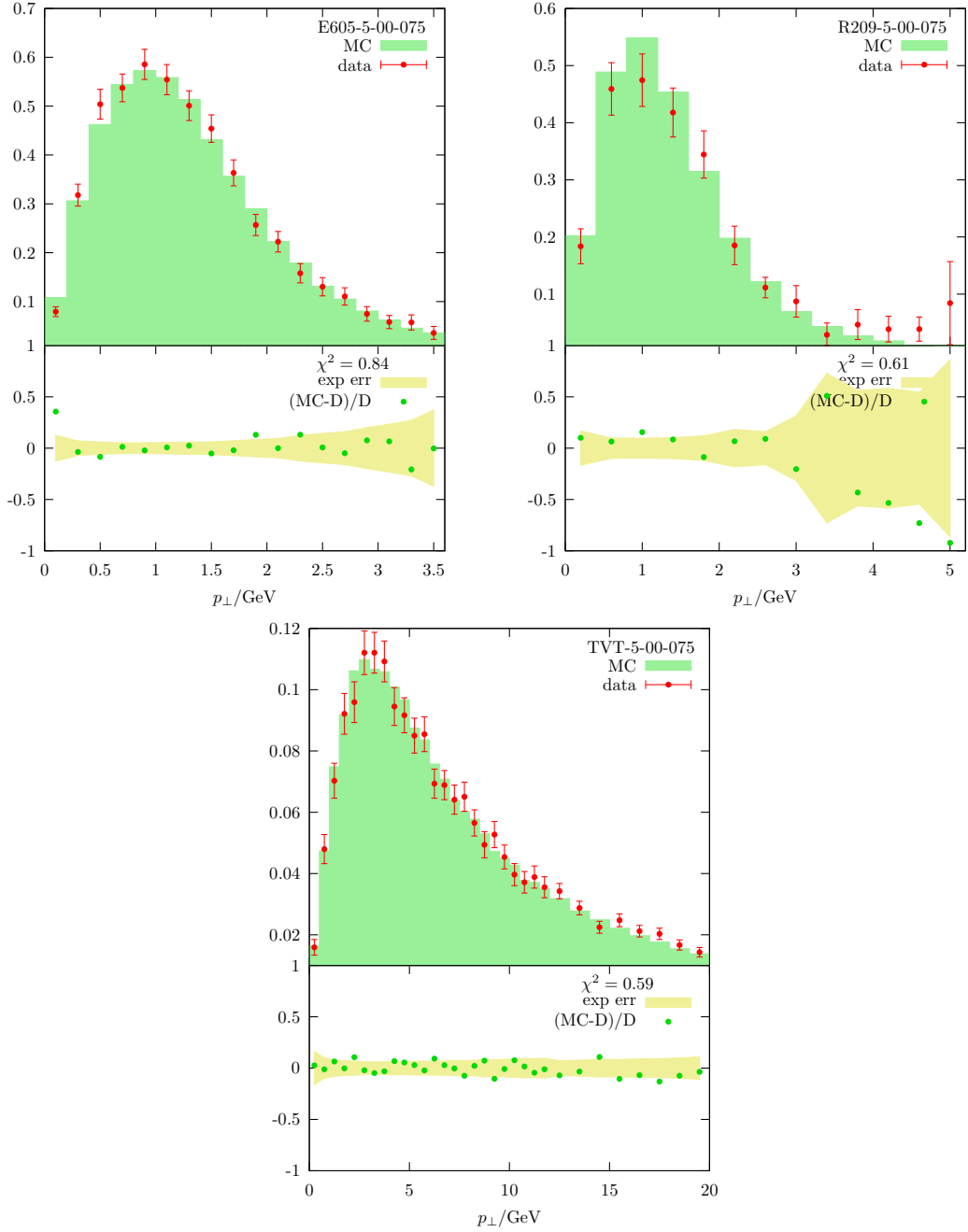
  \caption{Comparison of the parton level 
    results from the non--perturbative model
    with data from E605 (top left), R209 (top right) and CDF (bottom).
    The Monte Carlo results are from our parameter set with
    $\varphi_0=0.0, p_{\perp_0}=0.75\,$GeV.  Each panel includes two plots.  The
    upper plot compares MC to data directly, whereas the lower plot
    shows the ratio (MC-Data)/Data against the relative data error.  
    \label{fig:bestpt-parton} }
}
\FIGURE{%
  \includegraphics[scale=1.1]{ptZ.31}
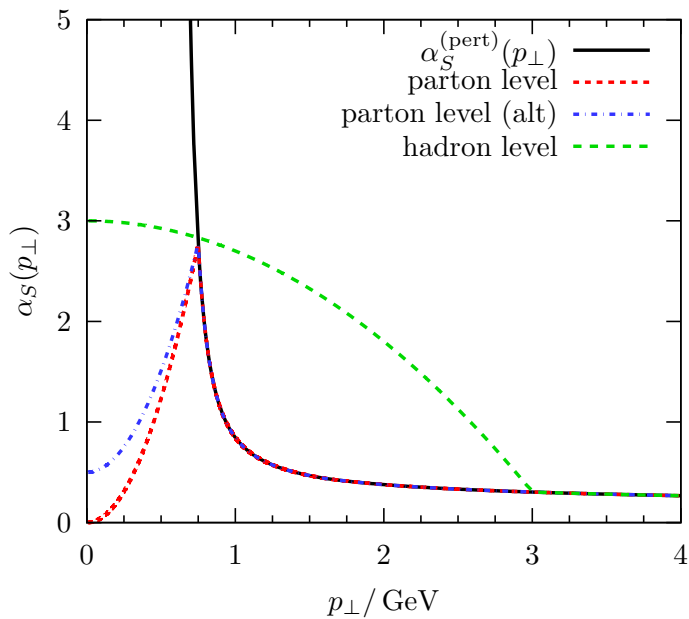
  \caption{ The optimal choice: ``\emph{quadratic}'' interpolation
    with $\alpha_S(0)=0$ and $p_{\perp_0} = 0.75\,$GeV is shown.  For
    comparison, we also show the purely perturbative $\alpha_S$ (LO)
    and another reasonable parametrisation of $\alpha_S$ in the
    non--perturbative region for our parton level results. In addition
    we show our best fit for the hadron level results.\label{fig:alpha_s} }
}

It is interesting to compare the $\alpha_s$ parametrisation we have
found with other approaches to modelling non-perturbative corrections
to inclusive observables with a modified coupling in the soft region
(see for example
Refs.~\cite{Dokshitzer:1995zt,Dokshitzer:1995qm,Guffanti:2000ep}).
Ref.~\cite{Dokshitzer:1995zt} finds an average value of the coupling
over the range from 0 to 2~GeV of about 0.5, while
Ref.~\cite{Dokshitzer:1995qm} argues that the effective coupling
should vanish at $p_\perp\to0$.  For our best-fit parametrisation, the
average value of the coupling over the range from 0 to 2~GeV is around
0.7.  Considering that their fits to data typically use NLO
calculations, while we have used a leading log parton shower, this
could be considered good agreement.  

\section{Hadron-level results}
\label{HadronLevel}

As we mentioned earlier, the results of the previous section are not
suitable for full event simulation, because the masslessness of the
light quarks and gluons is not consistent with the hadronization model
used in Herwig++.  Therefore in this section we perform the same
comparison with data but with the effective parton masses returned to
their default values, tuned to $e^+e^-$ annihilation data.

Performing an initial scan over parameter space we find that we need
to consider a much wider range of values than in the massless case.
We can get a good description of the data from each experiment, but
there is more tension between the three experiments leading to a
larger total $\chi^2$.  We choose as our best fit point
$\alpha_S(0)=3$ and $p_{\perp_0} = 3.0\,$GeV giving a $\chi^2$ per
degree of freedom of 0.94 for the Tevatron data and of 2.72 for all
data.  We show the results in Fig.~\ref{fig:bestpt-massive}.
\FIGURE[h]{%
  \includegraphics[scale=0.7]{ptscan.60}\hspace{1cm}
  \includegraphics[scale=0.7]{ptscan.61}\hfill\\[5pt]
  \includegraphics[scale=0.7]{ptscan.62}
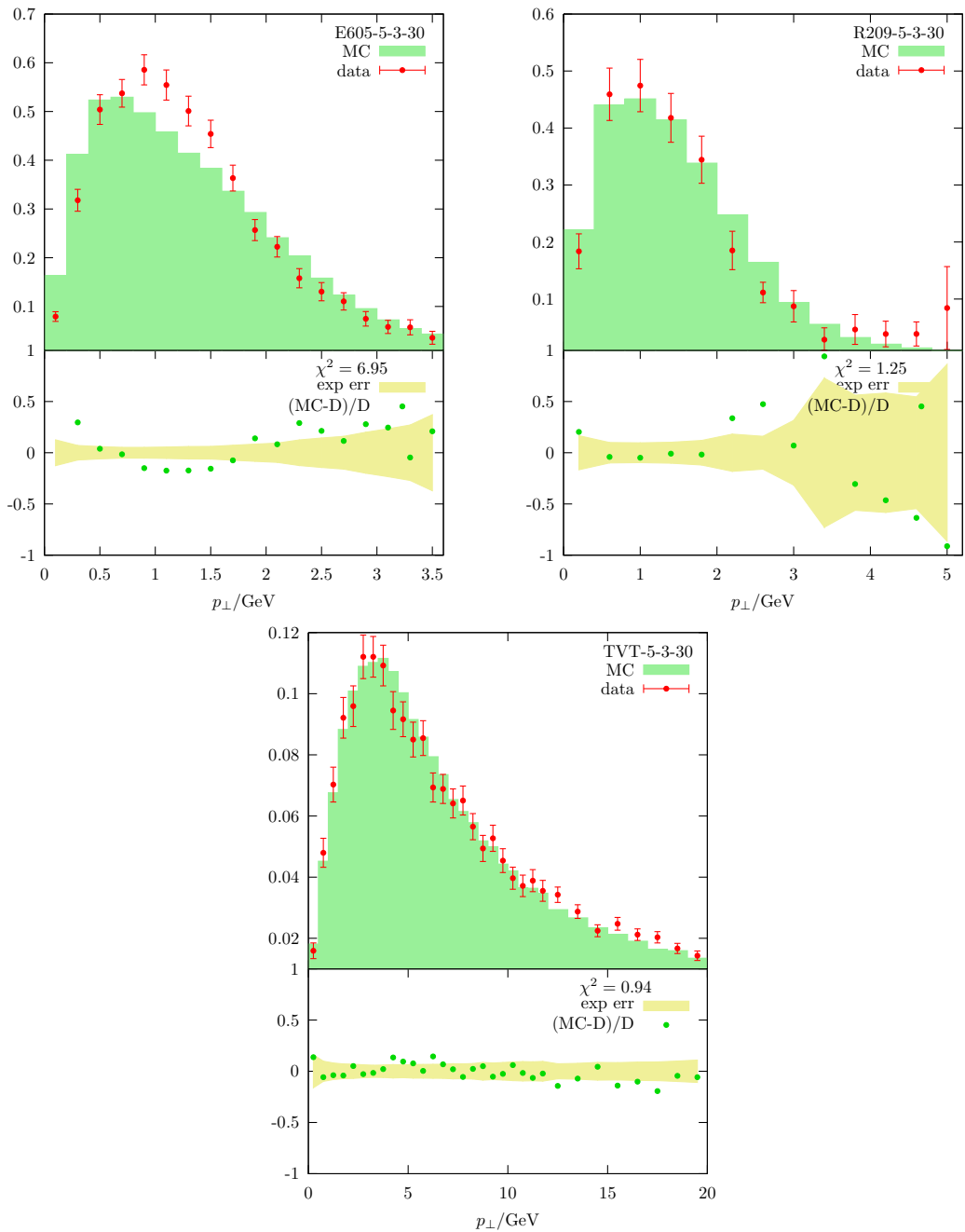
  \caption{As Fig.~\ref{fig:bestpt-parton} but for the combination of our
  non-perturbative emission model with the model of non-perturbative
  parton masses built in to Herwig++ by default.
   \label{fig:bestpt-massive} }
}

This time our best-fit $\alpha_S$ parametrization is very different
from those of
Refs.~\cite{Dokshitzer:1995zt,Dokshitzer:1995qm,Guffanti:2000ep}~-- it
is much larger in the non-perturbative region.  This is not surprising
since our coupling is now `fighting against' an emission distribution
that is already falling as $p_\perp\to0$ relative to the perturbative
one.  Although the overall description of data is somewhat worse with
the non-perturbative parton masses, it is acceptable, and we prefer to
maintain Herwig++'s description of final states so we keep this as our
default model for the remainder of the study.

\section{LHC result and comparison with other approaches}
\label{LHC}

\subsection[Z boson transverse momentum]{
  \boldmath Z boson transverse momentum}
In this section we would like to compare the result of extrapolating
our model to LHC energies with the results from two other approaches:
\ResBos{} \cite{Berge:2004nt} and Gaussian intrinsic $k_\perp$.   

First of all, we compare our prediction on the parton level (filled
histogram) and the hadron level (dot--dashed, blue).  Both histograms
give a consistent extrapolation.  We have tried different values of
$\alpha_S(0)$, ranging up to 1.5, for our parton level prediction and
find no visible effect.  This emphasises the relative unimportance of
the non--perturbative region for the description of this observable at
the LHC.  

The result from \ResBos{} in Fig.~\ref{fig:LHC-pt} (solid, black)
shows a slightly different behaviour from our prediction.  We predict
a slightly more prominent peak and a stronger suppression towards
larger transverse momenta.  The same trend is already visible when
comparing both approaches to Tevatron data although both are
compatible with the data within the given error band.  Both
computations match the data well at large transverse momenta as they
rely on the same hard matrix element contribution for single hard
gluon emission.  We want to stress the remarkable feature that we both
predict the same peak position with these models.  This is quite
understandable as both models are built on the same footing: extra
emission of soft gluons.  A comparison of \ResBos{} to data from
experiments at various energies was made in \cite{resbos-02}.
\FIGURE[t]{%
  \includegraphics[scale=1.1]{ptZ.30}
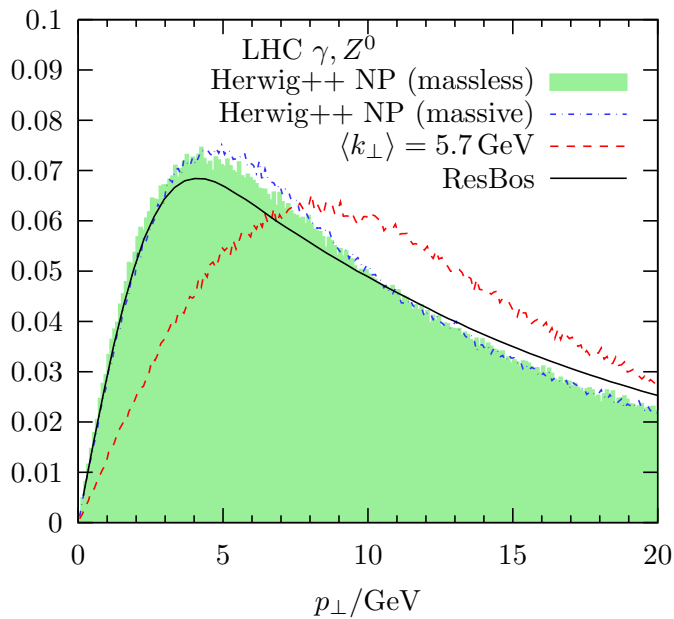
  \caption{Vector boson $p_\perp$ distribution at the LHC.  Our model
    is compared to the extrapolation of Gaussian intrinsic $k_\perp$
    to LHC energies and the result from \ResBos{}. 
    \label{fig:LHC-pt} }
}

Furthermore (dashed, red) we see the Herwig++ result from only using
intrinsic $\langle k_\perp\rangle = 5.7\,$GeV as recommended in
\cite{herwig++,herwigman}.  This large value stems from an extrapolation from
lower energy data with the assumption that the average $k_\perp$ will
depend linearly on $\ln(M/\sqrt{S})$.  The peak is seen to lie at a
considerably higher value of transverse momentum.  It would clearly be
of interest to have experimental data to distinguish these two models
of non--perturbative transverse momentum.

\subsection{Non-perturbative final state radiation}
\label{sec:fsr}

As briefly discussed in the introduction, we want to stress that the
approach of adding non--perturbative soft gluon radiation to the
parton shower should be connected to the non--perturbative input that
the parton shower is linked to in the initial state.  We think of this
radiation as originating from long--range correlations within the
coloured initial state.  

We have checked the effect of the same model for final state
radiation.  We find a dramatic increase in the amount of soft
radiation when we compare event shapes, simulated with our new model
for soft emissions, to LEP data, which are described well by the
default parton shower model.  Using the default hadronization model,
we observe a dramatic softening of the event shapes, leading to a poor
description of data.  However, the default hadronization model
produces a considerable amount of transverse momentum smearing during
cluster splitting and decay, and is tuned to data together with a
parton shower model that does not have non-perturbative smearing.
Therefore to turn on this smearing, without modifying, or at least
retuning, the hadronization model, must lead to a significant amount
of double-counting.  It is an interesting question, which we reserve
for future work, whether a good fit can be obtained with our model.

\section{Conclusion}

Aiming for a universal model of non--perturbative soft gluon radiation
we have achieved a reasonable description of data at three different
energies.  We consider the model based on soft gluon radiation,
much like the resummation program \ResBos{}, to have a more meaningful
physics input than simply extrapolating the Gaussian smearing of a
primordial transverse momentum. Of course, if this model is universal,
it should make predictions for other processes, such as jet and photon
production.  We plan to study these processes in more detail in the
future.

We also found that using our model as the only non-perturbative
ingredient in the simulation, i.e.\ removing the non-perturbative
constituent parton masses that usually cut off the parton shower in
Herwig++, gave a somewhat better description of the data.  This lays
open the speculation that perhaps, in some way, the two approaches
could be combined, using our model for initial-state radiation, and
the usual model, tuned to describe the final states of $e^+e^-$
annihilation, for final-state radiation.  We leave consideration of
this combination to future work however.

\acknowledgments We would like to thank the referee of the first
version of our paper for stimulating a more detailed study of our
results on the parton level.  We are grateful to Fred Olness and our
Herwig++ collaborators for fruitful discussions and to Stefan Berge
and Pavel Nadolsky for help with the \ResBos{} comparison. We
gratefully acknowledge support from the EU Marie Curie ESRT Fellowship
EUROTHEPHY, under contract number (MEST-CT-2005-020238) and EU Marie
Curie Research Training Network MCnet (MRTN-CT-2006-035606) which made
this project at CERN possible.  One of us (AS) also would like to
acknowledge the EU Marie Curie Research Training Network HEPTOOLS
(MRTN-CT-2006-035505) for additional support.

\appendix

\section{Herwig++ parameter settings}
\label{sec:parameters}
The study has been done with Herwig++ 2.2.0 \cite{herwig++,herwigman}.
%revision \texttt{svn-r2301}.
We ran with the default matrix element for $\gamma, {\rm Z}$
production with only initial state parton showers.  We left final
state parton showers and hadronic decays switched off as they were
irrelevant for this study.  The following parameters in release 2.2.0
are important to switch off the final state parton shower and to
adjust the intrinsic $p_\perp$:
\begin{verbatim}
  cd /Herwig/Shower
  set SplittingGenerator:FSR No
  set Evolver:IntrinsicPtGaussian 0.4*GeV
\end{verbatim}
Our preferred result, as shown in Fig.~\ref{fig:bestpt-massive}, was obtained
by setting 
\begin{verbatim}
  set AlphaQCD:NPAlphaS 5
  set AlphaQCD:Qmin 3.0*GeV 
  set AlphaQCD:AlphaMaxNP 3
\end{verbatim}
Here, ``\texttt{AlphaQCD:NPAlphaS 5}'' selects the quadratic non--perturbative
model.  The flat model would correspond to setting this parameter to
\texttt{6}. \texttt{AlphaQCD:Qmin} sets the value of $p_{\perp_0}$ and
\texttt{AlphaQCD:AlphaMaxNP} directly sets the value $\alpha_S(0)$.
As obtaining results for the parton level with very small masses and
cutoffs was very computing intensive, we also modified the code in
order to leave out the timelike showers from partons that have
been radiated in the initial state shower.

\end{document}